# IMAGE ENCRYPTION USING DIFFERENTIAL EVOLUTION APPROACH IN FREQUENCY DOMAIN

Ibrahim S I Abuhaiba[1] and Maaly A S Hassan[2]

[1]Department of Computer Engineering, Islamic University of Gaza, Gaza, Palestine
ihaiba@yahoo.com

[2]Department of Computer Engineering, Islamic University of Gaza, Gaza, Palestine
maaly_awad@hotmail.com

## ABSTRACT

*This paper presents a new effective method for image encryption which employs magnitude and phase manipulation using Differential Evolution (DE) approach. The novelty of this work lies in deploying the concept of keyed discrete Fourier transform (DFT) followed by DE operations for encryption purpose. To this end, a secret key is shared between both encryption and decryption sides. Firstly two dimensional (2-D) keyed discrete Fourier transform is carried out on the original image to be encrypted. Secondly crossover is performed between two components of the encrypted image, which are selected based on Linear Feedback Shift Register (LFSR) index generator. Similarly, keyed mutation is performed on the real parts of a certain components selected based on LFSR index generator. The LFSR index generator initializes it seed with the shared secret key to ensure the security of the resulting indices. The process shuffles the positions of image pixels. A new image encryption scheme based on the DE approach is developed which is composed with a simple diffusion mechanism. The deciphering process is an invertible process using the same key. The resulting encrypted image is found to be fully distorted, resulting in increasing the robustness of the proposed work. The simulation results validate the proposed image encryption scheme.*

## 1. INTRODUCTION

Owing to the advance in network technology, information security is an increasingly important problem. Popular application of multimedia technology and increasingly transmission ability of network gradually lead us to acquire information directly and clearly through images [1]. Hence, image security has become a critical and imperative issue [2]. Image encryption techniques try to convert an image to another image that is hard to understand; to keep the image confidential between users, in other word, it is essential that nobody could get to know the content without a key for decryption [3][4][5]. Furthermore, special and reliable security in storage and transmission of digital images is needed in many applications, such as pay-TV, medical imaging systems, military image communications and confidential video conferences, etc. In order to fulfill such a task, many image encryption methods have been proposed, but some of them have been known to be insecure [5], so we always in need to develop more and more secure image encryption techniques. Traditional data encryption techniques can be divided into two categories which are used individually or in combination in every cryptographic algorithm: substitution and transposition. In substitution technique, we symmetrically replace one symbol in the data with another symbol according to some algorithm; in a transposition technique, we reorder the position of symbols in the data according to some rule [6].

Image encryption approaches fall into two broad categories: spatial domain methods [7] and frequency domain methods [8]-[9]. The term spatial domain refers to the image plane itself, and approaches in this category are based on direct manipulation of pixels in an image. In these algorithms, the general encryption usually destroys the correlation among pixels and thus makes the encrypted images incompressible. Frequency domain processing techniques are based on modifying the Fourier transform of an image. The Fourier transform can be reconstructed (recovered) completely via an inverse process with no loss of information. This is one of the most important characteristics of this representation because it allows us to work in the "Fourier domain" and then return to the original domain without losing any information. Encryption techniques based on various combinations of methods from these two categories are not unusual [10]. In this paper we present a novel image encryption scheme which employs magnitude and phase manipulation using Differential Evolution (DE) approach. It deployed the concept of keyed discrete Fourier transform (DFT) followed by DE operations for encryption purpose. Firstly two dimensional (2-D) keyed discrete Fourier transform is carried out on the original image to be encrypted. Secondly crossover is performed between two components of the encrypted image, which are selected based on Linear Feedback Shift Register (LFSR) index generator. Similarly, keyed mutation is performed on the real parts of a certain components selected based on LFSR index generator. The LFSR index generator initializes it seed with the shared secret key to ensure the security of the resulting indices. The process shuffles the positions of image pixels. A new image encryption scheme based on the DE approach is developed which is composed with a simple diffusion mechanism. The deciphering process is an invertible process using the same key. The proposed method, dealing with private key cryptosystem, works in the frequency domain. The basis for the proposed method is that the encrypted image is obtained by magnitude and phase manipulation of the original image using the secret key. The original image magnitude and phase can be uniquely retrieved from the encrypted image if and only if the key is known. The resulting encrypted image is found to be fully distorted, resulting in increasing the robustness of the proposed work.

The remainder of this paper is organized as follows. In Section 2, we present some of the already existing related work. In Section 3, we talk about the properties of the Fourier transform. In Section 4, we present the method of differential evolution. In Section 5, we describe our proposed encryption method. We validate the proposed method through experiments in Section 6, and in Section 7 we discuss the results. Finally, we conclude in Section 8.

## 2. RELATED WORK

### A Technique for Image Encryption using Digital Signatures

A. Sinha and K. Singh [11] have proposed a new technique to encrypt an image for secure image transmission. The digital signature of the original image is embedded to the encoded version of the original image prior to transmission. Image encoding is done by using an appropriate error control code, such as a Bose-Chaudhuri Hochquenghem (BCH) code. At the receiver end, after the decryption of the image, the digital signature can be used to verify the authenticity of the image. This encryption technique provides three layers of security [12]. In the first step, an error control code is used which is determined in real-time, based on the size of the input image. Without the knowledge of the specific error control code, it is very difficult to obtain the original image. The dimension of the image also changes due to the added redundancy. This poses an additional difficulty to decrypt the image. Also, the digital signature is added to the encoded image in a specific manner. At the receiver end, the digital signature can be used to verify the authenticity of the transmitted image. The advantage of the scheme is the authenticity verification. Increment in the size of the image due to added redundancy is the disadvantage of the algorithm. Also it does not have any compression scheme.

### Lossless Image Compression and Encryption Using SCAN

S.S. Maniccam and N.G. Bourbakis [13] have presented a new methodology which performs both lossless compression and encryption of binary and gray-scale images. The compression and encryption schemes are based on SCAN patterns generated by the SCAN methodology. The SCAN is a formal language-based two-dimensional spatial-accessing methodology which can efficiently specify and generate a wide range of scanning paths or space filling curves. This algorithm has lossless image compression and encryption abilities. The distinct advantage of simultaneous lossless compression and strong encryption makes the methodology very useful in applications such as medical imaging, multimedia applications, and military applications. The drawback of the methodology is that compression-encryption takes longer time [12].

### A New Encryption Algorithm for Image Cryptosystems

C. C. Chang, M. S. Hwang, and T. S. Chen [14] use one of the popular image compression techniques, vector quantization to design an efficient cryptosystem for images. The scheme is based on vector quantization (VQ), cryptography, and other number theorems. In VQ, the images are first decomposed into vectors and then sequentially encoded vector by vector. Then traditional cryptosystems from commercial applications can be used. Major advantage of this algorithm [12], it has a simple hardware structure. Required bit rate of VQ is small. Since VQ compresses the original image into a set of indices in the codebook, we can save a lot of storage space and channel bandwidth. The other advantage is that VQ has a simple hardware structure for providing a fast decoding procedure.

### Color Image Encryption Using Double Random Phase Encoding

S. Zhang and M. A. Karim [15] have proposed a new method to encrypt color images using existing optical encryption systems for gray-scale images. The color images are converted to their indexed image formats before they are encoded. In the encoding subsystem, image is encoded to stationary white noise with two random phase masks, one in the input plane and the other in the Fourier plane. At the decryption end, the color images are recovered by converting the decrypted indexed images back to their RGB (Red-Green-Blue) formats. The proposed single-channel color image encryption method is more compact and robust than the multichannels methods. This technique introduces color information to optical encryption. An RGB color image is converted to an indexed image before it is encrypted using a typical optical security systems. At the decryption end, the recovered indexed image is converted back to the RGB image. Since only one channel is needed to encrypt color images, it reduces the complexity and increases the reliability of the corresponding optical color image encryption systems [12].

### An image encryption algorithm based on chaotic sequences

Yi Kai-Xiang and Sun Xing et al., [16] give an image encryption algorithm based on chaotic sequence. First, the real number value chaotic sequences using the key value is generated. Then it is dispersed in to symbol matrix and transformation matrix. Finally the image is encrypted using them in DCT domain. DCT is a lossy data compression technique, image may occur some distortions caused by lossy data compression and noise, but this method can still correctly decrypt and restore original image, and can achieve a high security degree.

## 3. PROPERTIES OF FOURIER TRANSFORMS: *Importance of phase and magnitude*

Equations indicate that the Fourier transform of an image can be complex. Both the magnitude and the phase functions are necessary for the complete reconstruction of an image from its Fourier transform. Fig. 1b shows what happens when Fig. 1a is restored solely on the basis of the magnitude information and Fig. 1c shows what happens when Fig. 1a is restored solely on the basis of the phase information, where |A| means the magnitude of image A and φ means its phase. Neither the magnitude information nor the phase information is sufficient to restore the image. The magnitude-only image (Fig. 1b) is unrecognizable and has severe dynamic range problems. The phase-only image (Fig. 1c) is barely recognizable, that is, severely degraded in quality.

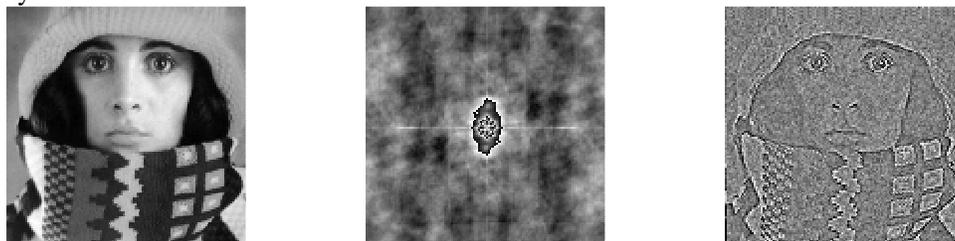

Fig. 1a Original image A [m,n]  Fig. 1b Restored image, φ = 0  Fig. 1c Restored image, |A| = constant

## 4. THE METHOD OF DIFFERENTIAL EVOLUTION

DE two main stages: crossover and mutation. The crossover procedure takes two selected vectors and combines them about a crossover point thereby creating two new vectors. The mutation procedure modifies a certain vector subject to a mutation function, introducing further changing into the original vectors [17]-[18].

### 4.1 DE Crossover and Mutation Example on Simple Data

#### 4.1.1 Crossover

Fig. 2 presents a trivial example that explains the method of DE: crossover and mutation operations. Applying DE crossover operation on a pair of binary vectors -as shown in Fig. 2a- will result in new pair of binary vectors. Firstly we divide each vector about a crossover point, thus creating two parts for each vector, and then we swap the second part of the first vector with the second part of the second vector.

[0001100010]              [0001100001]
[0000100001]              [0000100010]

[3   2]    crossover    [3   1]
[1   1]                 [1   2]

Fig. 2a DE crossover operation

#### 4.1.2 Mutation

Moreover, if we apply a certain mutation function on a single bit of a given binary vector we will get completely different vector. Suppose we define the mutation function to be NOT function (inverter), applying the inverter function on the forth bit of the original binary vector will result in a new different one as shown in Fig. 2b.

[0000100001]              [0000101001]
[1   1]    mutation    [1   9]

Fig. 2b DE mutation operation

# 5. THE PROPOSED ENCRYPTION METHOD

In this section we introduce our proposed image encryption scheme. Fig. 3 shows the general view of the main steps of our proposed DE frequency domain based cryptosystem.

The main idea is to firstly carry out the two dimensional (2-D) keyed discrete Fourier transform on the original image, resulting in the first level of image encryption by the use of the secret key. Secondly we perform crossover operation on two components of the encrypted image, which are selected based on Linear Feedback Shift Register (LFSR) index generator. Thus we make more shuffling to the positions of image pixels leading to fully distorted encrypted image. The LFSR index generator initializes it seed with the shared secret key value to ensure the security of the resulting indices. *At the third level*, we apply the keyed mutation function (real-part = SecretKey – real-part) on the real parts of a certain components selected similarly based on LFSR index generator. Our scheme is based on the DE approach which is composed with a simple diffusion mechanism. The basis for the proposed method is that the encrypted image is obtained by magnitude and phase manipulation of the original image using the secret key. The original image magnitude and phase can be uniquely retrieved from the encrypted image if and only if the key is known. This idea makes the cryptosystem secure.

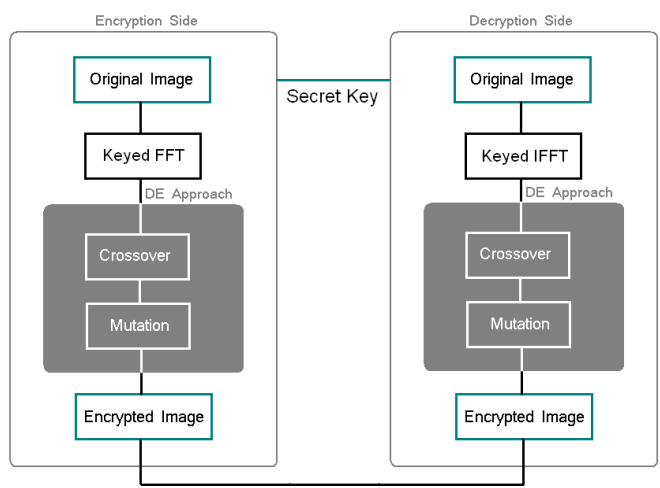

Fig. 3 Proposed cryptosystem general view

The proposed scheme has three major steps; Keyed discrete Fourier transform, Crossover operation and Mutation operation. The top-down approach as follow:

## 5.1 First Step: Computing the Keyed FFT of the Original Image

The Fourier transform is a representation of an image as a sum of complex exponentials of varying magnitudes, frequencies, and phases. It plays a critical role in   a broad range of image processing applications, including enhancement, analysis, restoration, and compression. Working with the Fourier transform on a computer usually involves a form of the transform known as the discrete Fourier transform (DFT). There are two principal reasons for using this form: 1) The input and output of the DFT are both discrete, which makes it convenient for computer manipulations, and 2) There is a fast algorithm for computing the DFT known as the fast Fourier transform (FFT).

The discrete Fourier transform of a function (image) f(x, y) of size $M \times N$ is given by the equation [5]

$$F(u,v) = \frac{1}{MN} \sum_{x=0}^{M-1} \sum_{y=0}^{N-1} f(x,y) e^{-j2\pi(\frac{ux}{M}+\frac{vy}{N})} \qquad (1)$$

This expression must be computed for values of u = 0, 1, 2, …, $M$-1, and also for v = 0, 1, 2, …, N-1. Similarly, given F(u ,v), we obtain the original function  f(x ,y) via the inverse Fourier transform, given by the expression [10]

$$f(u,v) = \sum_{u=0}^{M-1} \sum_{v=0}^{N-1} F(u,v) e^{j2\pi(\frac{ux}{M}+\frac{vy}{N})} \qquad (2)$$

for x = 0, 1, 2, …, $M$-1 and y = 0, 1, 2, …, $N$-1.

The $1/MN$ multiplier in front of the Fourier transform sometimes are placed in front of the inverse instead. Other times both equations are multiplied by $1/\sqrt{MN}$. The location of the multiplier does not matter. If two multipliers are used, the only requirement is that their product be equal to $1/MN$ [10].

In the proposed method, we sought to modify the traditional Fourier transform to become keyed Fourier transform.  The key ensures that applying the inverse Fourier transform does not give any meaningful result unless it was done by the use of the key.

Our approach to modify the traditional discrete Fourier transform as follow: we replaced the multiplier $\frac{1}{MN}$ in the traditional Fourier transform, equation (1), by the multiplier $\frac{1}{MN/Secret\ Key}$ as shown in the following equation (3)

$$F(u,v) = \frac{1}{MN/SecretKey} \sum_{x=0}^{M-1} \sum_{y=0}^{N-1} f(x,y) e^{-j2\pi(\frac{ux}{M}+\frac{vy}{N})} \qquad (3)$$

Similarly, we added the multiplier $\frac{1}{Secret\ Key}$ in front of the traditional inverse Fourier transform, equation (2), as shown in the following equation (4)

$$f(x,y) = \frac{1}{SecretKey} \sum_{u=0}^{M-1} \sum_{v=0}^{N-1} F(u,v) e^{j2\pi(\frac{ux}{M}+\frac{vy}{N})} \qquad (4)$$

Gain of this choice of multipliers, ensures that we maintain the restriction that their product is equal to $1/MN$.

The following example shows the first step of our proposed scheme, Fig. 4a shows the original data, Fig. 4b shows the effect of applying the keyed fast Fourier transform and Fig. 4c shows the reversibility of this step.

```
origimage =

        2            4          17         100
     3955            4          23         199
        1            3           4           5
        9            7           6           5
```

Fig. 4a Original data

```
keyedfft =

  1.0e+003 *

  4.3440              3.9170 + 0.2910i   3.6900              3.9170 - 0.2910i
  0.1100 - 4.1540i    0.1850 - 3.8350i  -0.0820 - 3.7720i   -0.2090 - 4.0230i
 -4.0720             -3.9530 - 0.0950i  -3.8660             -3.9530 + 0.0950i
  0.1100 + 4.1540i   -0.2090 + 4.0230i  -0.0820 + 3.7720i    0.1850 + 3.8350i
```

Fig. 4b Keyed fast Fourier transformed data

```
keyedifft =

        2            4          17         100
     3955            4          23         199
        1            3           4           5
        9            7           6           5
```

Fig. 4c Keyed inverse fast Fourier transformed data

## 5.2 Second Step: DE Crossover Operation

In this step, two frequency-domain components (complex numbers) are taken and combined about a crossover point thereby creating two new components. This crossovering could be achieved by treating the component as two parts: real and complex. In this case the crossover point is ready and all what we need to do is to swap the complex part of the first component with the complex part of the other, which will result in a change in the amplitude and the phase of the new components. In other words the resulting components will contain real and complex parts from different original component.

Suppose we have the following two original components:

$X=Real1+Complex1,$ $\qquad$ $Y=Real2+Complex2$

If they are crossovered as discussed above, the new generated components will be formed as follow:

$Z=Real1+Complex2,$ $\qquad$ $W=Real2+Complex1$

An important point arises here, how to select these components? These components are chosen based on the indices generated using LFSR index generator [19], leading to more and more diffusion. The idea here is to use the secret key as an initial seed of the LFSR index generator. In this research, secret key is 256-bits. Firstly, the secret key is divided into 32 segments each of them contains 8-bits. Each segment is assigned to 8-cells LFSR, which means that we have 32 8-cells LFSRs. Each 8-cell LFCR will result in a new 8-bits based on its feedback function. These 32 8-bits outputs will be concatenated to form 256-bits again. Finally, the 256-bits binary output will be converted into decimal value which represents the generated index. This new index will be considered as an initial seed to the LFSR index generator to generate another index. Indices generated in such fashion are the indices of our components to be crossovered.

To find row and column indices of two components to be crossovered, we firstly initialize the row index with SecretKey% NumberOfRows. Then we start *for loop*, the counter of the *for loop* is considered as a column index. The *for loop* performs two operations, firstly applies the crossover operation on the components located at (row index, *for loop* counter) and (row index, (*for loop* counter)+1) and secondly, updates the row index using LFSR random number generator, which takes the previous value of row index as an initial seed. The following Pseudo code clarifies the idea.

```
rowIndex=SecretKey%NumberOfRows
For i=1:2 … (NumberOfColumns-1)
        crossover(component(rowIndex,i),component(rowIndex,i+1))
        rowIndex=LFSR_RandomNumberGenerator(rowIndex)
End
```

```
crossoveredimage =

   1.0e+003 *

     4.3440 + 0.2910i   3.9170             3.6900 - 0.2910i   3.9170
     0.1100 - 3.8350i   0.1850 - 4.1540i  -0.0820 - 4.0230i  -0.2090 - 3.7720i
    -4.0720 - 0.0950i  -3.9530            -3.8660 + 0.0950i  -3.9530
     0.1100 + 4.0230i  -0.2090 + 4.1540i  -0.0820 + 3.8350i   0.1850 + 3.7720i
```

Fig. 5a Crossovered data

```
crossoveredcipher =

   1.0e+003 *

     4.3440             3.9170 + 0.2910i   3.6900             3.9170 - 0.2910i
     0.1100 - 4.1540i   0.1850 - 3.8350i  -0.0820 - 3.7720i  -0.2090 - 4.0230i
    -4.0720            -3.9530 - 0.0950i  -3.8660            -3.9530 + 0.0950i
     0.1100 + 4.1540i  -0.2090 + 4.0230i  -0.0820 + 3.7720i   0.1850 + 3.8350i
```

Fig. 5b Reversibility of crossover operation

In Fig. 5a, we continue with the previous example to explain the second step of our proposed scheme. Fig. 5a shows the results of crossovering the components in the first row with its corresponding components in the second row and the components of the third row with the components of the fourth row. Fig. 5b shows the reversibility of this operation.

This step is reversible. Both encryption and decryption sides use the same shared secret key as an initial seed to their LFSR index generator, therefore the same indices will be generated and consequently the crossover effect will be removed and the original components will be retrieved again.

## 5.3 Third Step: DE Mutation Operation

The mutation procedure modifies a certain components subject to a mutation function, introducing further changing into the components (amplitude and phase). Mutation will be applied on the components chosen based on the "index number generator" system Fig. 7 in a fashion as discussed in the crossover operation.

```
mutatedimage =

  1.0e+003 *

  -4.3390 + 0.2910i   3.9170              -3.6850 - 0.2910i   3.9170
  -0.1050 - 3.8350i   0.1850 - 4.1540i    0.0870 - 4.0230i   -0.2090 - 3.7720i
   4.0770 - 0.0950i  -3.9530               3.8710 + 0.0950i  -3.9530
  -0.1050 + 4.0230i  -0.2090 + 4.1540i    0.0870 + 3.8350i    0.1850 + 3.7720i
```

Fig. 6a Mutated data

```
mutatedcipher =

  1.0e+003 *

   4.3440 + 0.2910i   3.9170               3.6900 - 0.2910i   3.9170
   0.1100 - 3.8350i   0.1850 - 4.1540i    -0.0820 - 4.0230i  -0.2090 - 3.7720i
  -4.0720 - 0.0950i  -3.9530              -3.8660 + 0.0950i  -3.9530
   0.1100 + 4.0230i  -0.2090 + 4.1540i    -0.0820 + 3.8350i   0.1850 + 3.7720i
```

Fig. 6b Reversibility of mutation operation

Fig. 6a shows the results of mutating the components in the first and third columns using the following simple mutation function: realPart=5-realPart (this mutation function just for discussion purpose). Fig. 6b shows the reversibility of this operation using the same mutation function.

In our proposed method, we applied a keyed mutation function defined in equation (5); this keyed mutation function is **self invert able**

$$Mutation \ of \ F(u,v) = SecretKey - RealPart \ (F(u,v)) \qquad (5)$$

Mutation operation is also reversible, because of the same initial seed to the LFSR index generator -shared secret key- in both encryption and decryption sides. The mutation effect will be removed by applying the mutation function again on the same components selected based on the generated indices.

To find row and column indices of one component to be mutated, we firstly initialize the row index with SecretKey%NumberOfRows and the column index with (SecretKey-128)%NumberOfColumns. These initial indices will be of the first component to be mutated. Then we start **for loop** that performs two operations, firstly applies the mutation operation on the component located at (row index, column index) and secondly, updates the row and column indices using LFSR random number generator which takes the previous values of row and column indices as an initial seeds. The following Pseudo code shows the steps of the process.

```
rowIndex=SecretKey%NumberOfRows
columnIndex=(SecretKey-128)%NumberOfColumns
 For i=1 … (SizeOfImage*SecretKey)
      mutation(component(rowIndex,columnIndex))
      rowIndex=LFSR_RandomNumberGenerator(rowIndex)
      columnIndex=LFSR_RandomNumberGenerator(columnIndex)
End
```

We begin with different row and column indices in order to avoid restricting mutation operation on the diagonal elements (SecretKey,SecretKey), (newly generated number, newly generated number), ….. Since, similar seeds yield similar random numbers.

### 5.4 Random (index/number) generator

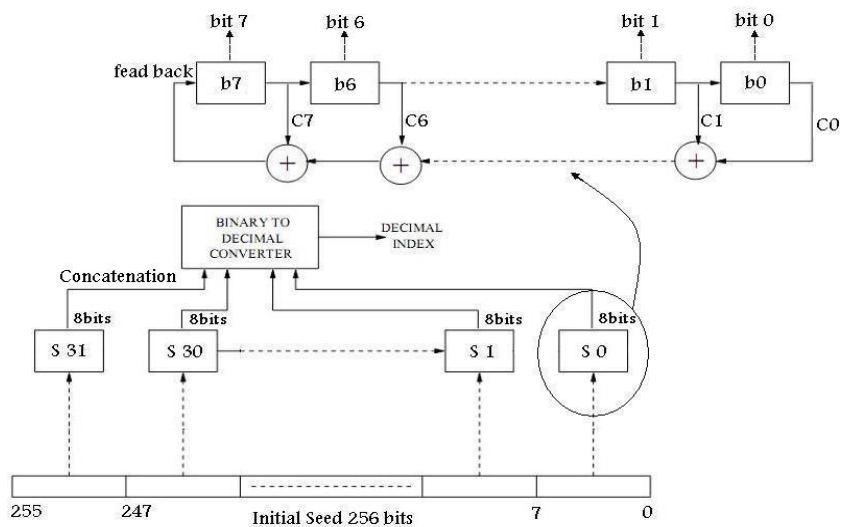

Fig. 7 LFSR block diagram [19]

Random index generator is used to generate the indices of the frequency domain components which will be encountered in each step of the proposed cryptosystem. In crossover operation, random number generator is used to generate row and column values of the components to be crossovered, likely it is used to generate the indices of rows and columns of the frequency domain components to be mutated. Fig. 7 illustrates the block diagram of the linear feedback shift registers (LFSR) index generator. The main idea behind the random number generator and how it works is shown below [19]:

*The random number generator consists of 32-LFSR named S0 to S31. Each LFSR has 8-cells.*

**Step 1: Initialization**
- Initialize the 32 8-cells shift registers S0 to S31 with 256-bits starting seed (transmitted through a secured channel as a shared secret key). The initial seed is divided into 32 partions each contains 8-bits, each of these partions are assigned to one corresponding 8-cells shift register. This idea is illustrated in the Fig. 8. The first partion contains bits from 0 to 7; this partion is assigned to 8-cells shift register S0.

**Step 2: Multiplication**
- For each 8-cells shift register (S0 to S31) multiply the outputs of the 8-cells with the coefficients (C7, C6, . . . . , C1, C0) of a primitive polynomial with respect to mod 2 operation, this primitive polynomial determines the feedback function of its shift register.
- Use this result as a feedback to the last cell.

**Step 3: Conversion and output**
- Convert the concatenated outputs of 32 8-cells LFSR from binary to decimal, this output will be 32x8 bits

  Output this number (this number represents the index)

To achieve the security in our proposed scheme, the initial seed is chosen to be the secret key which is shared between both encryption and decryption sides. Therefore, the choice of the indices of the components to be crossovered and mutated is based on the shared secret. Each shift register used in this paper consists of 8 cells (bits) as shown in Fig. 8. The following eighth degree primitive and irreducible polynomial is used. This polynomial is represented by the hexadecimal number: 11D, The corresponding binary form as follow: $(100011101)_2$

The primitive polynomial: $x8 + x4 + x3 + x2 + 1$

$b8 = b4 + b3 + b2 + b0$

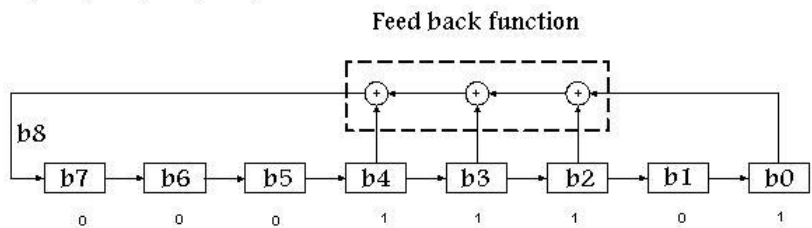

Fig. 8 eight cells LFSR index generator

## 6. EXPERIMENTAL RESULTS AND SECURITY ANALYSIS

To test our proposed encryption method, several experiments were performed. These measures were done on a laptop with microprocessor 1.83 GHz and Microsoft Windows XP platform. Programs are written using MATLAB version 7.0.1 and were applied on well known images, the behavior of our cryptosystem through encryption and decryption phase as shown bellow. Results of some experiments are given to prove its efficiency of application to digital images.

### 6.1 Validation of the proposed scheme

#### 6.1.1 Experiment I:

We used the gray-scale Cameraman image of size 256x256 and coloured Forest image of size 447x301 as the original images (plainimages). The encrypted (cipherimages) images are depicted in Fig. 9(b)-10(b). As shown, the regions of the encrypted images are totally invisible. We note that our cryptosystem can be adopted to encrypt grayscale and color images. The decryption method takes as input the encrypted image (cipherimage), together with the same secret key. The decrypted images are shown in Fig. 9(c)-10(c). The visual inspection of Fig. (9-10) shows the possibility of applying the proposed DE approach in frequency domain successfully in both encryption and decryption. Also, it reveals its effectiveness in hiding its contained information.

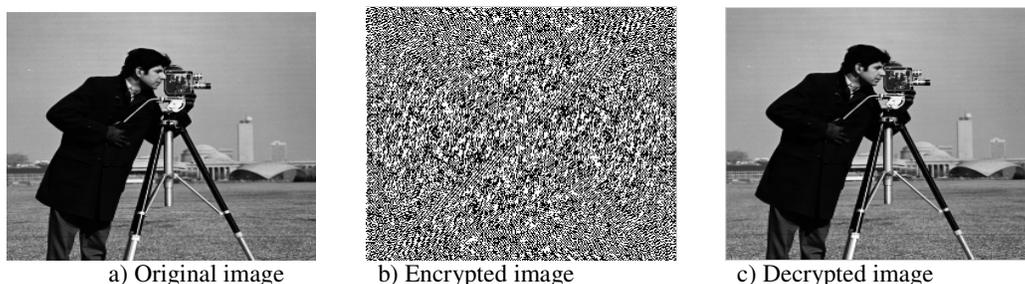

a) Original image      b) Encrypted image      c) Decrypted image

Fig. 9 The response of DE cryptosystem on a gray scale Cameraman Plainimage/Cipherimage

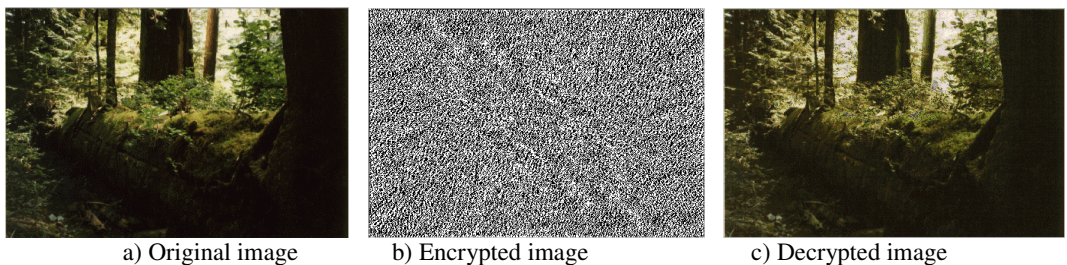

<div align="center">a) Original image    b) Encrypted image    c) Decrypted image</div>
<div align="center">Fig. 10 The response of DE cryptosystem on a coloured Forest Plainimage/Cipherimage</div>

### 6.1.2 Experiment II:

In the following experiment we apply the proposed DE method on a mesh text image; system response is shown in Fig. 11.

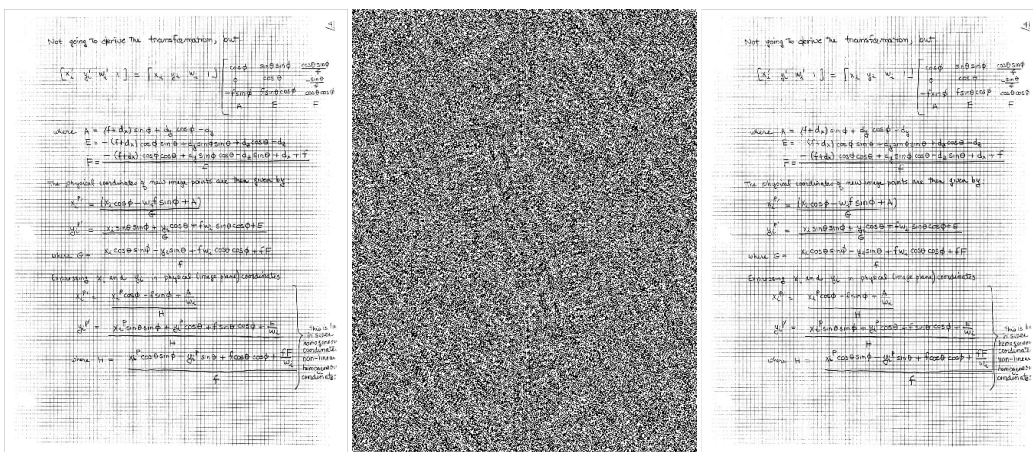

<div align="center">a) Original image    b) Encrypted image    c) Decrypted image</div>
<div align="center">Fig. 11 The response of DE cryptosystem on a mesh text image</div>

The encrypted (cipherimage) image is depicted in Fig. 11(b). As shown, the regions of the encrypted image are totally invisible. The decrypted image is shown in Fig. 11(c). From these Figures, it is clearly noticeable that the proposed DE cryptosystem respond very well on a mesh text image which can be considered as a complicated kind of images. Also, it reveals its effectiveness in hiding its contained information.

## 6.2 Security Analysis

A good encryption procedure should be robust against all kinds of cryptanalytic, statistical and brute force attacks. In this sub section, we discuss the security analysis of the proposed method such as key space analysis, statistical analysis, and sensitivity analysis with respect to the key to prove that the proposed cryptosystem is secure against the most common attacks [20]-[5].

### 6.2.1 Key space analysis

For a secure image cryptosystem, the key space should be large enough to make the brute force attack infeasible. The proposed method has 2256 different combinations of the secret key. An image cipher with such a long key space is sufficient for reliable practical use. In the proposed method, LFSR index generator is employed. The seed of the LFSR index generator is 256 bits; this seed is initialized with the secret key value.

### 6.2.2 Statistical analysis

It is well known that many ciphers have been successfully analyzed with the help of statistical analysis and several statistical attacks have been devised on them. Therefore, an ideal cipher should be robust against any statistical attack. To prove the robustness of the proposed scheme, we have performed statistical analysis by calculating the histograms and the correlations of two adjacent pixels in the plainimage/cipherimage.

### A. Histograms analysis

To prevent the leakage of information to an opponent, it is also advantageous if the cipherimage bears little or no statistical similarity to the plainimage. The histograms of the original images Fig. 13(b)-14(b) illustrates how pixels in the original images are distributed by graphing the number of pixels at each gray level. We have calculated and analyzed the histograms of the several encrypted images as well as its original images that have widely different content.

One typical example among them is shown in Fig. 12(b)-13(b). The histograms of a plainimages contain large spikes. These spikes correspond to gray scale values that appear more often in the plainimage. The histograms of the cipherimages as shown in Fig. 12(d)-13(d), are significantly different from that of the original images, and bear no statistical resemblance to the plainimages. It is clear that the histogram of the encrypted image are significantly different from the respective histogram of the original image and hence does not provide any clue to employ any statistical attack on the proposed image encryption procedure.

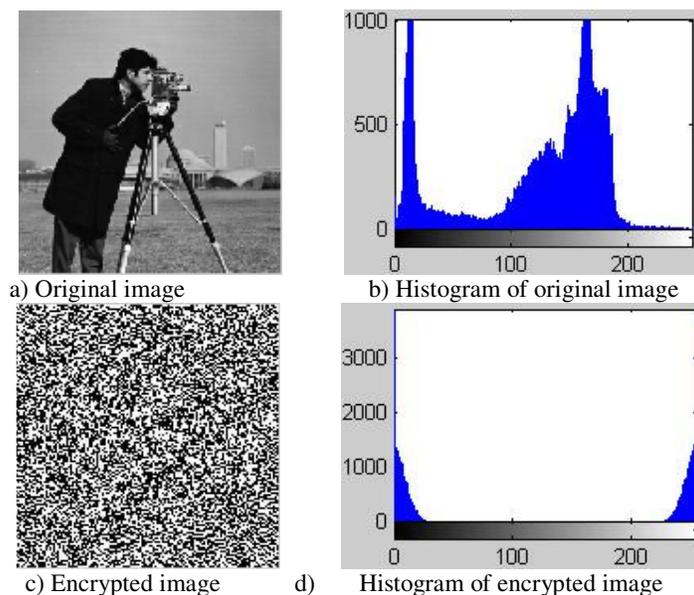

a) Original image      b) Histogram of original image

c) Encrypted image      d)    Histogram of encrypted image

Fig. 12 Histograms of the Cameraman plainimage and the corresponding cipherimage

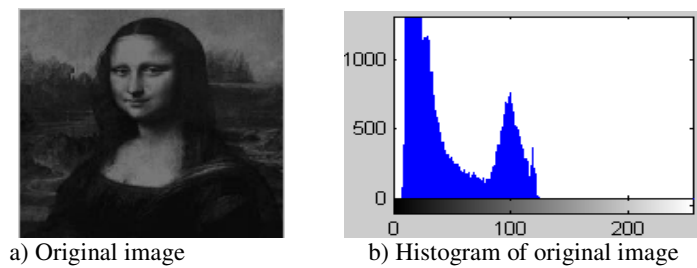

a) Original image      b) Histogram of original image

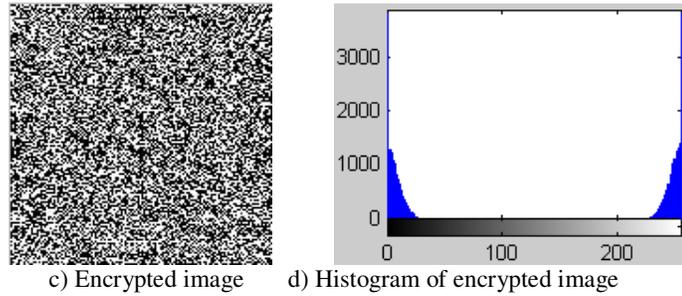

c) Encrypted image    d) Histogram of encrypted image

Fig. 13 Histograms of the Monaliza plainimage and the corresponding cipherimage

### B. Correlation coefficient analysis

In addition to the histogram analysis, we have also analyzed the correlation between two vertically adjacent pixels, two horizontally adjacent pixels and two diagonally adjacent pixels in plainimage/cipherimage respectively. Firstly, we randomly select 2000 pairs of two adjacent pixels from an image. Then, we calculate their correlation coefficient using the following two formulas [21]:

$$cov(x, y) = E(x - E(x))(y - E(y)), \qquad (6)$$

$$r_{xy} = \frac{cov(x, y)}{\sqrt{D(x)}\sqrt{D(y)}}, \qquad (7)$$

Where $x$ and $y$ are the values of two adjacent pixels in the image. In numerical computations, the following discrete formulas were used [21]:

$$E(x) = \frac{1}{N}\sum_{i=1}^{N} x_i \qquad (8)$$

$$D(x) = \frac{1}{N}\sum_{i=1}^{N} (x_i - E(x))^2, \qquad (9)$$

$$cov(x, y) = \frac{1}{N}\sum_{i=1}^{N} (x_i - E(x))(y - E(y_i)), \qquad (10)$$

Table 1 Correlation coefficients in plainimage/cipherimage

| Direction of adjacent pixels | Plainimage | Cipherimage |
|---|---|---|
| Horizontal | 0.9898 | 0.0303 |
| Vertical | 0.9805 | 0.0302 |
| Diagonal | 0.9769 | 0.0311 |

The correlation coefficients of two horizontally adjacent pixels are 0.9898 and 0.0303 respectively for both plainimage/cipherimage of our proposed method. Similar results for vertical and diagonal directions are obtained as shown in Table 1. It is clear from Table 1 that there is negligible correlation between the two adjacent pixels in the cipherimage. However, the two adjacent pixels in the plainimage are highly correlated.

### 6.2.3 Key sensitivity analysis

An ideal image encryption procedure should be sensitive with respect to the secret key. The change of a single bit in the secret key should produce a completely different encrypted image, which means that the cipherimage cannot be decrypted correctly although there is only a slight difference between encryption and decryption keys. This guarantees the security of the proposed method against brute-force attacks to some extent. For testing the key sensitivity of the proposed image encryption method, we have performed the following steps:

An original image in Fig. 14a is encrypted by using the secret key "1551917990046475381" which is equivalent to "1589853085422475" (in hexadecimal) and the resultant image is referred as encrypted image A as shown in Fig. 14b.

The same original image is encrypted by making the slight modification in the secret key i.e. "2704839494653322357" which is equivalent to "2589853085422475" (in hexadecimal) (the most significant bit is changed in the secret key) and the resultant image is referred as encrypted image B as shown in Fig. 14c.

Again, the same original image is encrypted by making the slight modification in the secret key i.e. "1551917990046475380" which is equivalent to "1589853085422474" (in hexadecimal) (the least significant bit is changed in the secret key) and the resultant image is referred as encrypted image C as shown in Fig. 14d. Finally, the three encrypted images A, B and C are compared.

In Fig. 14, we have shown the original image as well as the three encrypted images produced in the aforesaid steps. It is not easy to compare the encrypted images by simply observing these images. So for comparison, we have calculated the correlation between the corresponding pixels of the three encrypted images. For this calculation, we have used the same formula as given in equation (7) except that in this case x and y are the values of corresponding pixels in the two encrypted images to be compared. In Table 2, we have given the results of the correlation coefficients between the corresponding pixels of the three encrypted images A, B and C. It is clear from the table that no correlation exists among three encrypted images even though these have been produced by using slightly different secret keys. Key sensitivity analysis shows that changing one bit in encryption key will result in a completely different cipherimage.

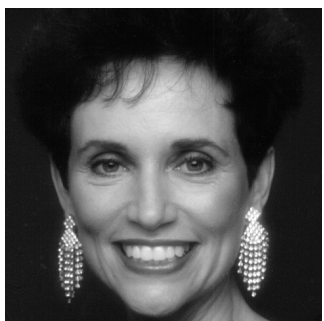

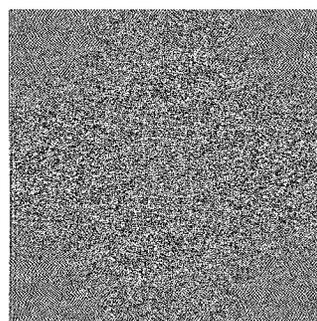

14a Original image         14b Encrypted image A with key
"1589853085422475"

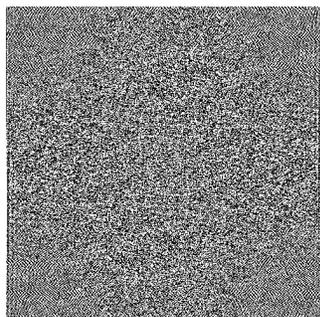 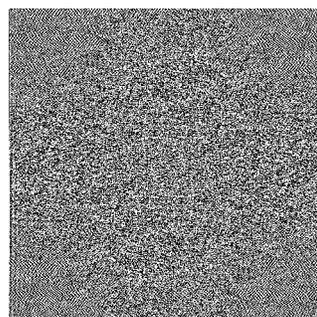

| 14c Encrypted image B with key | 14d Encrypted image C with key |
|---|---|
| "2589853085422475" | "1589853085422474" |

Fig. 14 Key sensitive test result 1 with the proposed scheme

Table 2 Correlation coefficients between the corresponding pixels of the three different encrypted images obtained by using slightly different secret key of an image shown in Fig. 14.

| Image 1 | Image 2 | Correlation coefficient |
|---|---|---|
| Encrypted image A Fig. 14b | Encrypted image B Fig. 14c | 0.0309 |
| Encrypted image B Fig. 14c | Encrypted image C Fig. 14d | 0.0358 |
| Encrypted image C Fig. 14d | Encrypted image A Fig. 14b | 0.0342 |

Moreover, in Fig. 15, we have shown the results of some attempts to decrypt an encrypted image with slightly different secret keys than the one used for the encryption of the original image. Particularly, in Fig. 15a and Fig. 15b respectively, the original image and the encrypted image produced using the secret key "1589853085422475" (in hexadecimal) are shown whereas in Fig. 15c and Fig. 15d respectively, the images after the decryption of the encrypted image (shown in Fig. 15b) with the secret keys "1589853085422475" (in hexadecimal) and "1589853085422474" (in hexadecimal). It is clear that the decryption with a slightly different key fails completely and hence the proposed image encryption procedure is highly key sensitive.

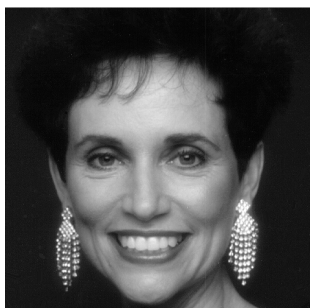 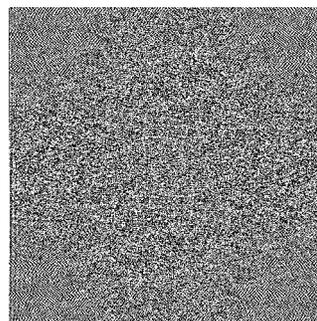

| 15a Original image | 15b Encrypted image with key |
|---|---|
| | "1589853085422475" |

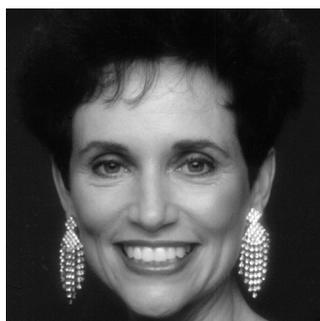 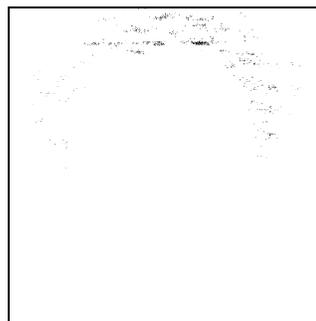

15c Decrypted image with key "1589853085422475"   165 Decrypted image with key "15898530854224754"

Fig. 15 Key sensitive test result 2 with the proposed scheme

### 6.2.4 Lossless and Opacity

Clearly from the above experiments' results, we can note that the proposed encryption method was lossless, since the decrypted images is exactly similar to the original images without any loss of data through encryption and decryption operations of this method, which means that there is no recorded noise in the decrypted images. We can also note that the opacity between the original images and the encrypted images is very high. On other words, the distortion between the original and encrypted images as shown in the above experiments is very high.

### 6.2.5 Complexity

To measure the complexity of the proposed encryption method, the time in seconds for doing the encryption and decryption operations for the above experiments was recorded in Table 3.

Table 3 Enciphering/deciphering speed test results of the proposed DE approach

|  | Size of data | Encryption Operation | Decryption Operation |
|---|---|---|---|
| Cameraman Image (256 x 256) | 63.7KB | 2.547000 seconds | 2.453000 seconds |
| Monaliza Image (256 x 256) | 27.1KB | 2.515000 seconds | 2.453000 seconds |
| Woman image (500 x 500) | 244KB | 9.687000 seconds | 9.437000 seconds |
| Lena image (512 x 512) | 49.9KB | 9.937000 seconds | 9.875000 seconds |
| Forest image (447 x 301) | 121KB | 5.235000 seconds | 5.110000 seconds |

## 7. CONCLUSION AND FINAL THOUGHTS

In the digital world nowadays, the security of digital images become more and more important since the communications of digital products over open network occur more and more frequently. In this paper, a new way of image encryption scheme have been proposed which utilizes the main steps of Differential Evolution (DE) approach: crossover and mutation on the frequency domain components of the plainimage, therefore changing these components' amplitude and phase in order to achieve more confusion and diffusion in the cipherimage. We have carried out key space analysis, statistical analysis, and key sensitivity analysis to demonstrate the security of the new image encryption procedure. According to the results of our security analysis, we conclude that the proposed m is expected to be useful for real-time image encryption and transmission applications. The future research in will be expanded to apply this method on multimedia data. Some image compression technique may also be included.

**Authors**

Prof. Ibrahim Abuhaiba is a professor at the Islamic University of Gaza, Computer Engineering Department. He obtained his Master of Philosophy and Doctorate of Philosophy from Britain in the field of document understanding and pattern recognition. His research interests include computer vision, image processing, document analysis and understanding, pattern recognition, artificial intelligence.

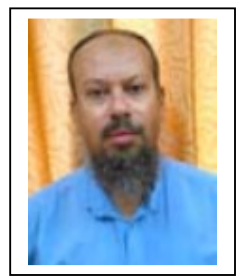

Eng. Maaly Hassan is pursuing her Master degree in Computer Engineering at the Islamic University of Gaza. She obtained her Bachelor degree in Computer Engineering from IUG, her area of interest include Image Processing, Multimedia and Mobile Ad Hoc and Sensor Networks.

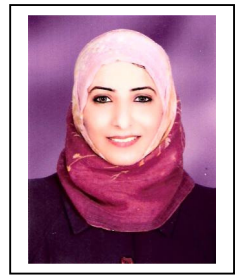